\documentclass[aps, prb, graphicx, 12pt, reprint, superscriptaddress, citeautoscript]{revtex4-1}
\usepackage[utf8]{inputenc}

\usepackage{graphicx}
\usepackage{amsmath}
\usepackage{amssymb}
\usepackage[hidelinks]{hyperref}
\usepackage{multirow}
\usepackage{soul}
\usepackage{xcolor}
\usepackage{cancel}

\usepackage[font=small,labelfont=bf]{caption} 
\usepackage[version=3]{mhchem} 

\newcommand{\e}{\ensuremath{\varepsilon}}

\newcommand{\nvirgin}{\ensuremath{N_\mathrm{S}}}

\newcommand{\ntotal}{\ensuremath{N_\mathrm{total}}}
\newcommand{\ki}{\ensuremath{\beta}}
\newcommand{\kd}{\ensuremath{\gamma}}

\newcommand{\cvirgin}{\ensuremath{S}}

\newcommand{\cactive}{\ensuremath{I}}

\newcommand{\cdone}{\ensuremath{R}}

\newcommand{\sir}{SIR}
\newcommand{\kt}{\ensuremath{\bar{\e}}}
\newcommand{\kto}{\ensuremath{\bar{\e}_0}}

\newcommand{\dist}{\ensuremath{n_\mathrm{S}(\e, t)}}

\newcommand{\disto}{\ensuremath{n_\mathrm{S}(\e,0)}}

\newcommand{\rdist}{\ensuremath{\hat{r}_\mathrm{I}(\e,t)}}

\newcommand{\ri}{\ensuremath{r_\mathrm{I}}}
\newcommand{\rr}{\ensuremath{r_\mathrm{R}}}
\newcommand{\hit}{\ensuremath{p_\mathrm{C}}}
\newcommand{\Ro}{\ensuremath{\mathcal{R}_0}}

\newcommand{\power}{\ensuremath{p}}

\newcommand{\emax}{\ensuremath{\e_\mathrm{max}}}

\newcommand{\order}{\ensuremath{{\mathcal{K}(t)}}}
\newcommand{\progress}{\ensuremath{\phi(t)}}

\begin{document}

\title{Heterogeneity in susceptibility dictates the order of epidemiological models}

\author{Christopher Rose}
\affiliation{School of Engineering, Brown University, Providence, Rhode Island, 02912, USA}

\author{Andrew J. Medford}
\affiliation{School of Chemical \& Biomolecular Engineering, Georgia Institute of Technology, Atlanta, Georgia, 30332, USA}

\author{C. Franklin Goldsmith}
\affiliation{School of Engineering, Brown University, Providence, Rhode Island, 02912, USA}

\author{Tejs Vegge}
\affiliation{Department of Energy Conversion and Storage, Technical University of Denmark, 2800 Kgs.~Lyngby, Denmark}

\author{Joshua S. Weitz}
\email{jsweitz@gatech.edu}
\affiliation{School of Biological Sciences, Georgia Institute of Technology, Atlanta, Georgia, 30332, USA}

\author{Andrew A. Peterson}
\email{andrew\_peterson@brown.edu}
\affiliation{School of Engineering, Brown University, Providence, Rhode Island, 02912, USA}
\affiliation{Department of Energy Conversion and Storage, Technical University of Denmark, 2800 Kgs.~Lyngby, Denmark}

\begin{abstract}
\noindent
The fundamental models of epidemiology describe the progression of an infectious disease through a population using compartmentalized differential equations, but do not incorporate population-level heterogeneity in infection susceptibility.
We show that variation strongly influences the rate of infection, while the infection process simultaneously sculpts the susceptibility distribution.
These joint dynamics influence the force of infection and are, in turn, influenced by the shape of the initial variability.
Intriguingly, we find that certain susceptibility distributions (the exponential and the gamma) are unchanged through the course of the outbreak, and lead naturally to power-law behavior in the force of infection; other distributions often tend towards these ``eigen-distributions'' through the process of contagion.
The power-law behavior fundamentally alters predictions of the long-term infection rate, and suggests that first-order epidemic models that are parameterized in the exponential-like phase may systematically and significantly over-estimate the final severity of the outbreak.
\end{abstract}

\maketitle

Mathematical models of disease dynamics divide a population into categories based on infection status; \textit{e.g.}, susceptible (S), infectious (I), and recovered/removed (R).
In the basic \sir\ model, the dynamics of individuals in each compartment can be written as~\cite{Bjornstad2020}
\begin{align}
\frac{d \cvirgin}{dt} &= - \ri \label{eq:ODEs} \\
\frac{d \cactive}{dt} &= \phantom{-} \ri  - \rr \nonumber,
\end{align}

\noindent
such that $\cvirgin + \cactive + \cdone = 1$.
The Kermack--McKendrick~\cite{Kermack1927} formulation---the basis for conventional, modern epidemiology models---assumes rates of infection and recovery to be $\ri = \ki \cactive \cvirgin$ and $\rr = \kd \cactive$, respectively, with \ki\ and \kd\ taken as rate constants with dimensions of inverse time.
These simplified models provide epidemiologists and policymakers with valuable intuition on the progression of an outbreak, and form the basis for more complex models that include such effects as geography, travel, latency, susceptibility to re-infection, stochasticity, and vital dynamics~\cite{Anderson1992,Keeling2007,Bertozzi2020,Moghadas2020}.
Previous studies have explored nonlinear forms for the rate of infection, \ri; many forms have been suggested including power laws and sub-exponential growth~\cite{Wilson1945,Liu1986,Liu1987,Hethcote1991,Regoes2002,Chowell2016}.

The \sir\ model assumes homogeneity of risk, an assumption unlikely to hold in practice.
That is, a real population will have a distribution of susceptibilities which can be based on a mixture of behavioral attributes (such as the number of people encountered in a typical day or interaction modalities) and inherent attributes (such as age, immune status, genetic differences, or varied responses to vaccines)~\cite{Longini1996,Halloran1996,Woolhouse1997,Dwyer1997,Dwyer2000,Smith2005,Izhar2015,King2018}.
Here, we confine our analysis to the simplest case where individuals have a static susceptibility, noting that dynamic changes in behavior can be accounted for separately~\cite{Eksin2019}.
While measurement of susceptibilities is certainly not straightforward, most studies suggest that a small percentage of the population carries a majority of the population's total susceptibility~\cite{Corder2020,Dwyer1997,King2018,Smith2005,Endo2020}.
If there is variation, then individuals that are more susceptible should tend to be infected earlier, leading to changes in the susceptibility distribution.
As a result, the most susceptible individuals will be disproportionately removed from the pool at the early stages of an outbreak, so that not just the \emph{number} of people in the susceptible pool will decrease, but also the \emph{average susceptibility} of the pool will decrease, both of which should slow the rate of spread of infections. 

One way to incorporate variability into the \sir\ model is to directly account for susceptibility in the rate equations.
Other epidemic models have incorporated such variability by explicitly accounting for assortative mixing \cite{Britton2020} or variation in contacts \cite{Dufresne2020}, by implicitly structuring a population based on variation in susceptibility \cite{Gomes2020}, or by the use of network-based models~\cite{Bansal2007}.
These analyses make different assumptions about the link between variation in risk and infection dynamics.
Despite the use of different assumptions, all of these models suggest that variation in the risk of transmission can lower the herd immunity threshold (\textit{i.e.}, the fraction of the population that must become immune for the disease to decrease in prevalence) when compared to predictions from equivalent homogeneous mixing models.
This agreement suggests it may be possible to develop a unified framework to understand the joint dynamics of infection and susceptibility.

\begin{figure*}
\centering
\includegraphics[width=0.99\textwidth]{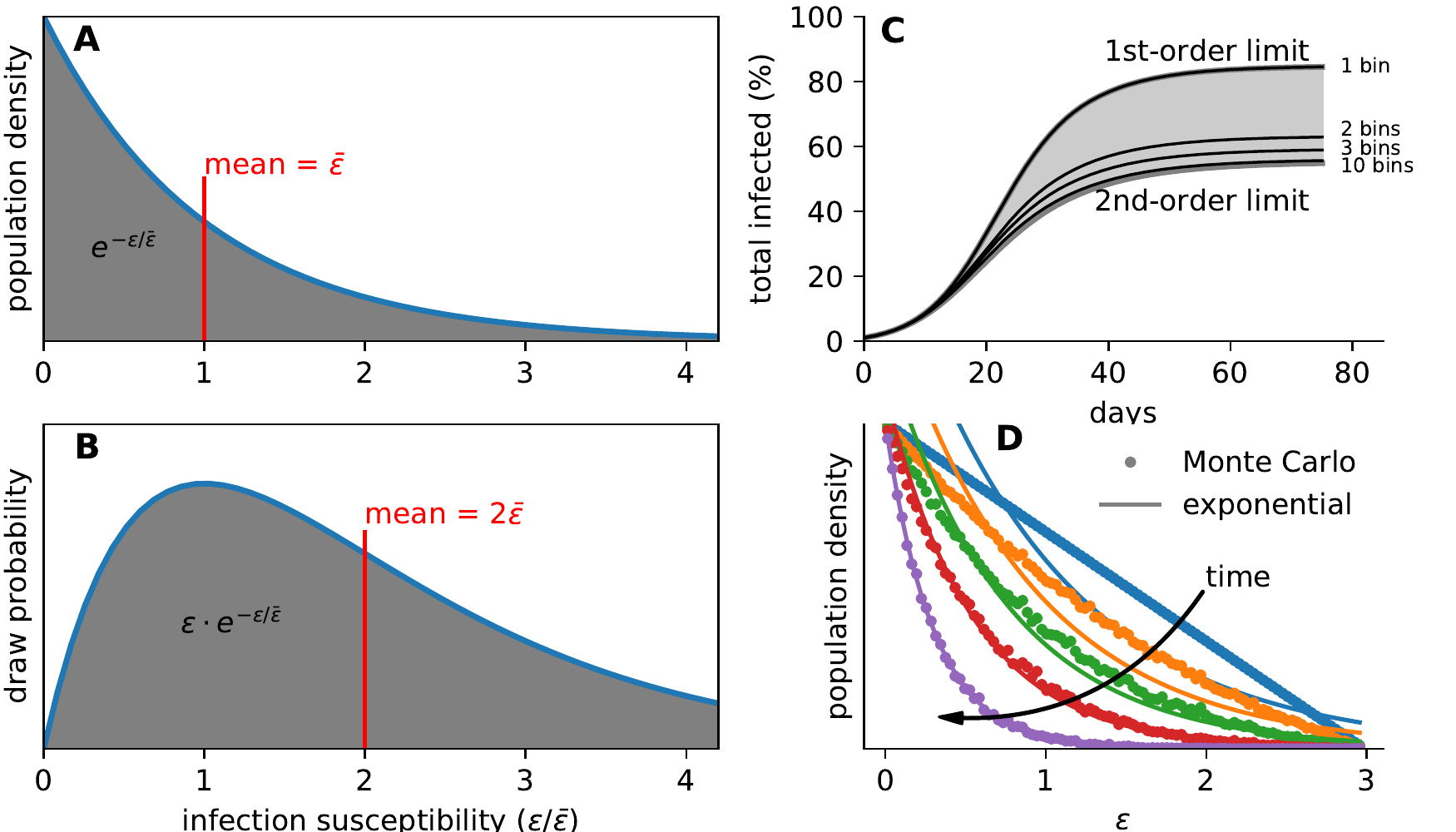}
\caption{ \textbf{A} and \textbf{B} show a continuous distribution of infection susceptibility in a population and the probability of the next individual infected, respectively. For an exponential distribution, the mean of the lower curve is double the mean of the upper curve. \textbf{C} compares numerically integrating a 1st- and 2nd-order \sir\ model with that obtained from binning the susceptibility into discrete levels consistent with an initially-exponential distribution. \textbf{D} shows that an exponential distribution emerges  when stochastically sampling from an initially linear distribution. The solid lines are exponential curves with the same mean as the Monte Carlo points of the same color.
\label{fig:susceptibility}}
\end{figure*}

Here, we define an individual's infection susceptibility, \e, such that individuals with \e=2 will become infected at twice the rate, on average, as individuals with \e=1, given they exist in the same population of infected people.
This definition is generic, and captures heterogeneity due to behavioral and/or inherent factors, such as those listed earlier.
If the distribution of susceptibilities in a population is given by \dist, shown for an exponential distribution in Figure~\ref{fig:susceptibility}A; then the infection rate is distributed as
\begin{equation}\label{eq:diff-rate}\rdist = \ki \cactive \e \frac{\dist}{\ntotal} \propto \e \cdot \dist \text{ ,} \end{equation}
\noindent
where \ntotal\ is the total number of individuals in the population.
This distribution, which we refer to as the ``draw probability'' ---that is, the conditional probability that a just-infected individual has susceptibility \e---is shown in Figure~\ref{fig:susceptibility}B, where it is apparent that the mean of the draw probability is higher than the mean of the susceptible pool.
This will be true of \emph{any} (non-singular) population distribution, putting downward pressure on the mean.
The overall rate of new infections at any time is then
\begin{align}
\ri &= \int_0^\infty \ki \, \cactive \, \e \, \frac{\dist}{\ntotal} \, d \e \nonumber \\
\ri &= \ki \, \cactive \, \kt \, \cvirgin \hspace{2em} \text{(any distribution)}, \label{eq:epsilonSIR}
\end{align}
\noindent
where $\kt(t)$ is the mean susceptibility of the pool, since $\kt(t) = \frac{1}{\nvirgin} \int_0^\infty \e \, \dist \, d\e$ with \nvirgin\ as the total number of susceptible people in the population.
This rate form ($\ki \kt \cactive \cvirgin$) is general; for example, if the entire population has an identical susceptibility of \e=1, \eqref{eq:epsilonSIR} simplifies to the classic \sir\ rate.

In principle, the only constraints on values of \e\ in \dist\ are that $\e \ge 0$ and that \kt\ is finite; a natural starting point is to assume that susceptibility follows the maximum-entropy distribution (\textit{i.e.}, the least-informative default) under these constraints, which for positive values with a specified mean is the exponential distribution~\cite{Jaynes_2003}
\[ \frac{\dist}{\nvirgin} = \frac{e^{-\e/\kt(t) }}{\kt(t)}.  \]
For this distribution, the draw probability has a mean value of $2 \, \kt(t)$ (at any time, provided the distribution remains exponential); that is, the average individual who has just gotten infected was twice as susceptible as the susceptible population as a whole.
(All derivations and assumptions not explicitly detailed in this manuscript are contained in the SI.)
We can then express how the pool's average susceptibility \kt\ is affected by the removal of high-\e\ individuals.
Since, on average, the susceptibility of a newly infected individual is $2\kt$, the pool's total susceptibility decreases as $d(\kt \nvirgin)/d\nvirgin = 2 \kt$.
Upon integration, we find the \emph{average} susceptibility of the pool decreases in direct proportion to the number of individuals left in the pool:
\begin{equation}\label{eq:kt-decrease} \kt(t) = \cvirgin(t), \end{equation}
\noindent
(where, without loss of generality, we define the initial population to have $\kt = 1$).
As a result, the rate of new infections (from \eqref{eq:epsilonSIR}) becomes second order in $S$,
\[  \ri = \ki \cactive \cvirgin^2 \hspace{2em} \text{(exponential dist.).} \]

A similar result should be observed by ``binning'' the susceptibility into discrete levels and introducing a separate ODE for each susceptibility level, as has been employed in previous literature~\cite{Langwig2017,King2018}; the match should improve as the number of bins increases, becoming exact at infinite binning.
We show the results of numerically integrating such a binned system given an initially exponential susceptibility distribution in Figure~\ref{fig:susceptibility}C.
As the number of discrete susceptibility levels is increased, we approach the second-order behavior predicted by this analysis.

The above analysis holds if the distribution is expected to remain exponential throughout the course of the outbreak.
Since a distribution will evolve as $\partial \dist / \partial t = - \ki \, \cactive \, \e \, \dist$, then the time-course of the distribution will follow

\begin{equation}\label{eq:evolution} \dist = \disto \cdot e^{- \e \progress}, \end{equation}

\noindent
where \disto\ is the initial distribution and $\progress \equiv \ki \int_0^t \cactive \, dt$ is a dimensionless progress variable that is monotonic with time.
\progress\ can be thought of as the cumulative infectious driving force.
Therefore, an initially exponential distribution will evolve as

\[ \dist = \frac{\ntotal}{\kto} \exp \left\{ - \left( \frac{1}{\kto} + \progress \right) \cdot \e \right\}, \]

\noindent
where \kto\ is the initial mean.
That is, the distribution stays exponential with respect to \e\ at all times, validating the continued use of the exponential distribution in \eqref{eq:kt-decrease}.

Interestingly, we also find that many starting distributions evolve toward an exponential form under the action of contagion.
This is shown for a linearly-decreasing distribution in Figure~\ref{fig:susceptibility}D, which we observed via Monte Carlo simulations to tend towards an exponential, through a stochastic sampling process consistent with the draw probability.
Indeed, this sculpting behavior can be inferred from \eqref{eq:evolution}, which for an initially linearly-decreasing distribution is proportional to $(\emax - \e) \cdot \exp \left\{ - \progress \e \right\}$, where the exponential term becomes dominant as \progress\ increases.

\begin{figure*}
\centering
\includegraphics[width=1.0\textwidth]{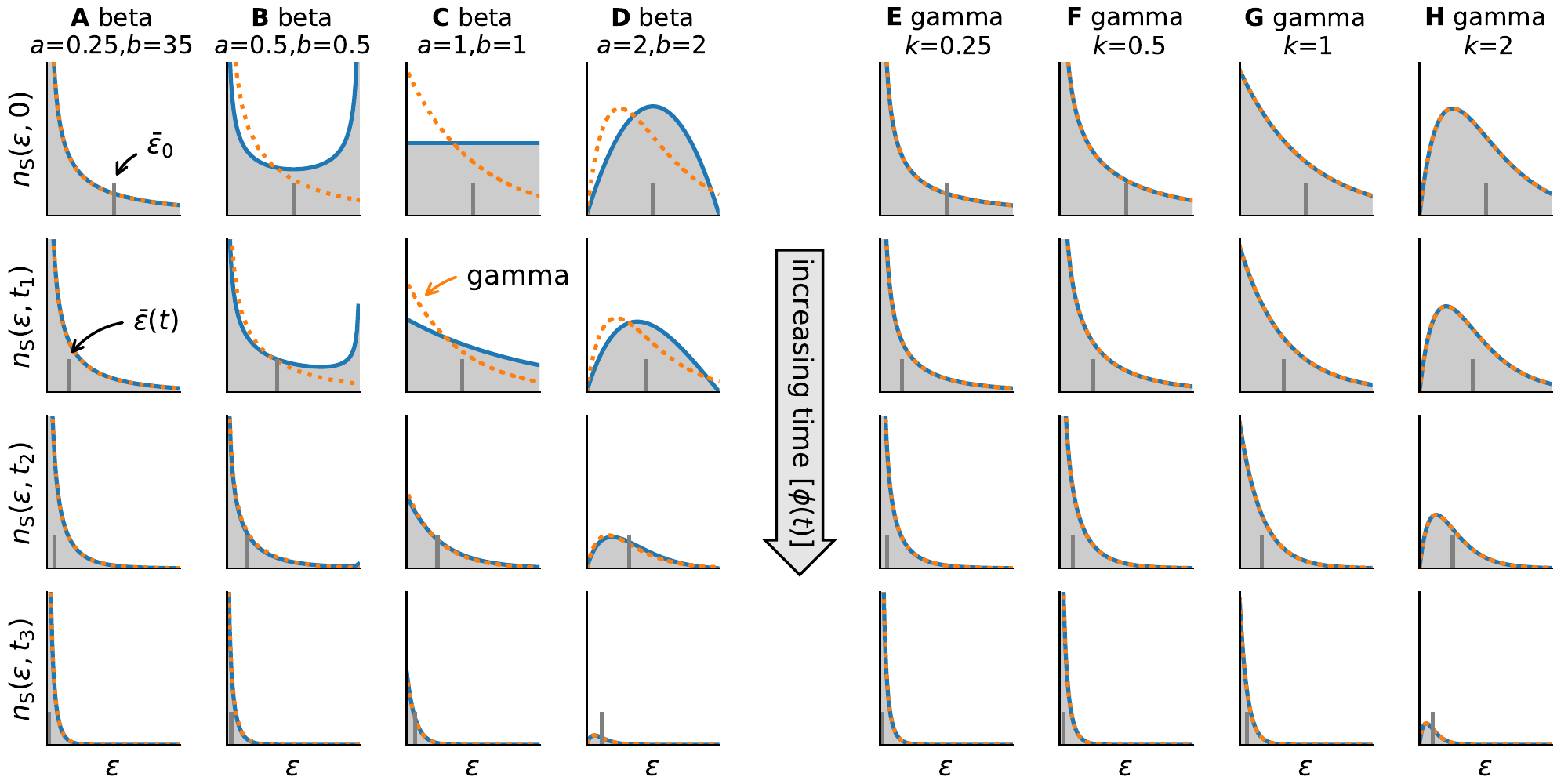}
\caption{The sculpting of susceptibility distributions under the act of contagion; that is, by \eqref{eq:evolution}. Each column represents the initial distribution indicated at the top of that column, with the shape of that distribution changing with increasing time (tracked by \progress) proceeding down the column; at values of \progress = 0, $\frac{1}{2}$, 2, and 4. In A--D, the orange dashed line shows a gamma function with the same mean and shape parameter $k=a$; in E--H the orange line is included to emphasize that a gamma distribution retains its shape. The small vertical bars indicate \kt, which is set to 1 in all $t=0$ plots. All plots are scaled identically; that is, all share the same limits on both the abscissa and ordinate. \label{fig:distribution-evolution}}
\end{figure*}

We can use \eqref{eq:evolution} to describe how contagion sculpts any arbitrary susceptibility distribution.
For convenience, we employ the beta distribution, which allows us to craft distributions of characteristic initial shapes, and in Figure~\ref{fig:distribution-evolution} we use this distribution to create initial distributions that are longer-tailed~(A), bimodal~(B), uniform~(C), and unimodal~(D).
Here, we can clearly see how the process of contagion sculpts distributions; features with large \e\ are quickly diminished; that is, the individuals with high susceptibility to infection are preferentially removed from \dist\ at early times.
In these sculpting plots, we also show a gamma distribution for reference, with its shape parameter $k$ equal to the $a$ parameter of the beta distribution, and we see that the contagion process sculpts these distributions towards their gamma counterparts.
Parts E--H of this figure show the sculpting of gamma distributions, which we see retain their shape during the act of contagion.
Indeed, it can be derived that any well-behaved initial distribution---specifically one that is continuous and can be expressed as a power series---will approach a gamma distribution under the sculpting process of \eqref{eq:evolution}; we show this in the SI, which also links the beta's $a$ parameter to the gamma's $k$ parameter.
After a more detailed analysis of the gamma distribution, we will show how the initial distribution shape and the sculpting process interact with the outbreak dynamics.

\begin{figure*}
\centering
\includegraphics[width=1.0\textwidth]{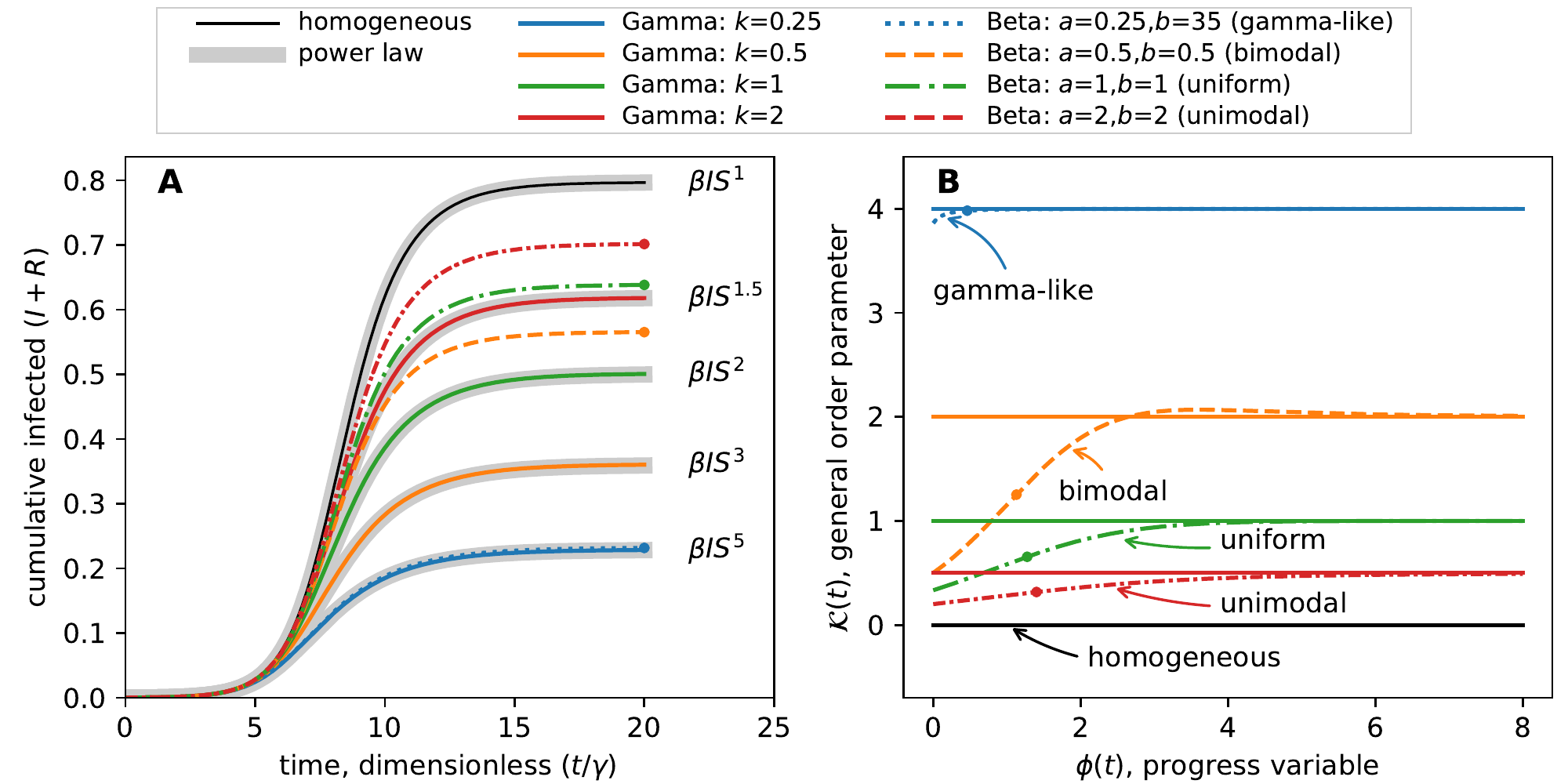}
	\caption{ How the initial susceptibility distribution affects the joint dynamics of infection and distribution sculpting. \textbf{A} Infection dynamics for the distributions shown in Figure~\ref{fig:distribution-evolution}. The gray background curves show power-law models ($\ki \cactive \cvirgin^\power$) for comparisons. Time is non-dimensionalized by the recovery rate constant. \textbf{B} Effective order parameter \order\ for the same distributions. For the gamma distribution, the order is always $1/k$; for the beta distributions, the order approaches that of a gamma distribution as the sculpting process proceeds. The dot on each beta curve indicates the value at $t/\kd=20$. The legend applies to both \textbf{A} and \textbf{B}. \label{fig:general-order}}
\end{figure*}

The gamma distribution is often used to describe variations in susceptibility~\cite{Dwyer1997,Smith2005,Langwig2017,King2018}.
This distribution, 
\[ \frac{\disto}{\ntotal} =  \frac{\e^{k-1} e^{-\e k / \kto} }{ \left( \frac{\kto}{k} \right)^k \Gamma(k) }, \]
is controlled via a shape parameter $k$ and simplifies to the exponential when $k{=}1$.
If $k{<}1$, the gamma becomes a longer-tailed distribution (such as an ``80:20'' distribution, where 80\%\ of the variation is captured by the most susceptible 20\%\ of the population), and as $k{\to}\infty$ the distribution becomes a Dirac delta function.
In published studies, $k$ is typically less than one~\cite{Dwyer1997,Smith2005,Gomes2020,Corder2020}.
As we observed in Figure~\ref{fig:distribution-evolution} and formalize in the SI, the gamma distribution is a general solution to the differential equation governing the dynamic sampling process; this makes it an ``eigen-distribution'' of the force of infection.
That is, if a population is initially gamma-distributed it will remain gamma-distributed---with the same shape parameter $k$ and decreasing mean---under the action of \eqref{eq:evolution}.

It can be derived that the mean susceptibility, for any given distribution, changes as
\[ \frac{d\kt}{d\cvirgin} = \frac{\sigma^2}{\kt \, \cvirgin} \hspace{2em} \text{(any distribution)} \]
where $\sigma^2(t)$ is the variance of the distribution.
Since the variance of the gamma function is $\sigma^2(t) = \kt(t)^2/k$ and $k$ stays constant during contagion, it follows that the mean susceptibility scales as
\[ \kt = \cvirgin^{1/k} \hspace{2em} \text{(gamma dist.)} \]
for an initial outbreak, giving a rate
\[ \ri = \ki \cactive \cvirgin^{1 + \frac{1}{k}} \hspace{2em} \text{(gamma dist.)}. \]
\noindent
This is exact power-law behavior for gamma-distributed susceptibilities, where the power is given by $\power = 1 + 1/k$.
(Power-law behavior has been suggested elsewhere in models of behavioral change~\cite{Eksin2019}; here, the behavior emerges naturally from the distribution.)
In the case of exponentially-distributed susceptibility ($k{=}1$), second-order behavior emerges.
A longer tail, corresponding to small values of $k$, can significantly increase the power; for example, the ``80:20'' distribution ($k{\approx}0.25$) pushes the order to approximately five.
It is only in the unlikely limit that all individuals are identical ($k{\to}\infty$) that we recover the first-order power of the classic \sir\ model, suggesting the standard \sir\ model relies on an extreme assumption about variation in susceptibility. 
This power law provides a convenient framework for capturing the effects of heterogeneous susceptibility through a single additional power-law parameter, which is dictated by the variability of susceptibility.

\begin{figure*}
\centering
\includegraphics[width=0.99\textwidth]{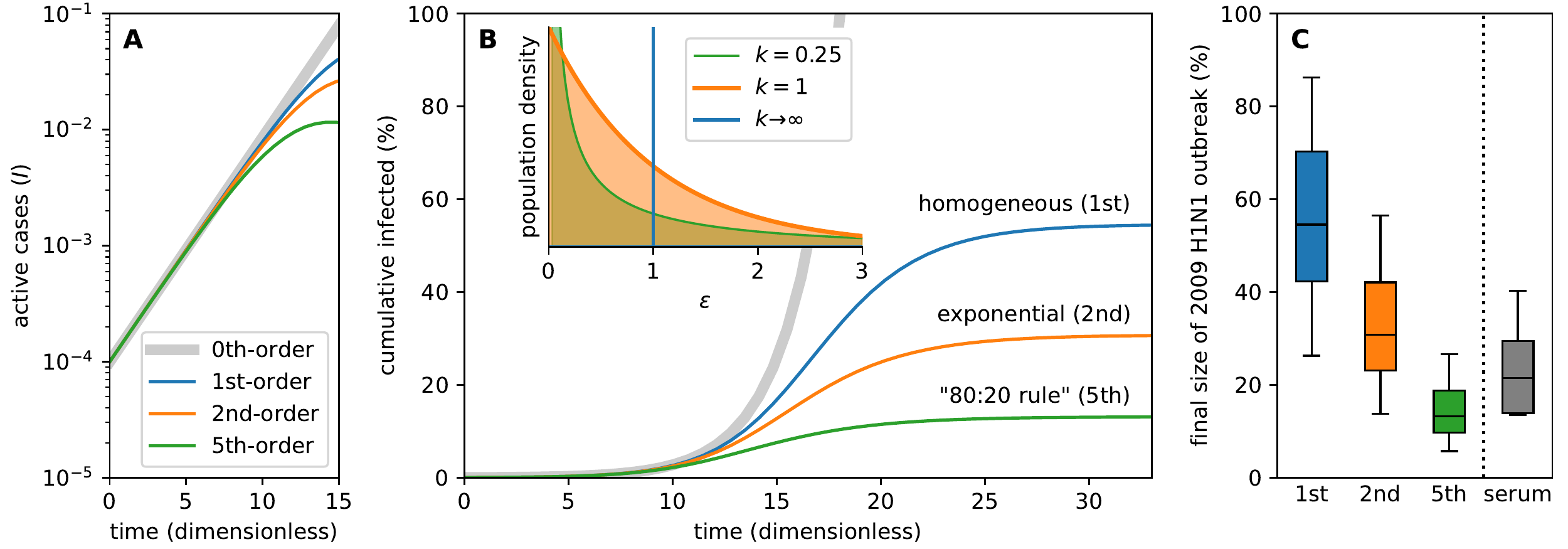}
\caption{
Outbreak dynamics and model comparison to the 2009 H1N1 outbreak.
\textbf{A}: The higher-order models are indistinguishable from one another in the early, exponential-like region (marked as ``0th-order''), where the models are frequently parameterized.
\textbf{B}:
Outbreak progression with various assumed susceptibility distributions.
The inset shows a gamma distribution at different $k$ values.
In all models, $\ki/\kd (= \Ro)$ is set to 1.47, the median value estimated from 78 studies at the early stages of the 2009 H1N1 outbreak,~\cite{Biggerstaff2014} and time is non-dimensionalized by the recovery rate ($\kd \cdot t$).
\textbf{C}: Final size predictions based on the 78 $R_0$ values measured~\cite{Biggerstaff2014} at early stages of the infection, in the 1st-, and 2nd-, and 5th-order models, compared with 11 seroepidemiological measurements obtained~\cite{Nishiura2011} after the outbreak stabilized.\label{fig:h1n1}}
\end{figure*}

Next we show how the sculpting process and the outbreak dynamics interact, allowing us to understand how the initial shape of the susceptibility distribution affects the dynamics of disease spread.
Conventionally, when outbreak models (\textit{e.g.}, the \sir)  are analyzed they are numerically integrated, and when differences in susceptibility are desired this can be achieved by ``binning'' the susceptibility into many discrete histogram bins~\cite{Langwig2017,King2018}, as we did in Figure~\ref{fig:susceptibility}C.
However, it is straightforward to track the cumulative progress of the epidemic \progress\ during the numerical integration.
Insofar as we can analytically express the mean susceptibility time course $\kt(\progress)$ for a given initial distribution under the action of \eqref{eq:evolution}, we can model outbreak dynamics for arbitrary initial distributions without resorting to the computational expense or complexity involved with binning. 
We give full details of how this can be simply implemented in the SI.
We note that methods have been introduced in prior literature to integrate such a system without binning, notably a method that approximates the partial differential equation as a truncated set of ordinary differential equations involving the moments of the distribution, making subsequent assumptions about relationships between these moments~\cite{Dushoff1996,Dwyer1997,Dushoff1999}.
In contrast, the method here gives an exact solution, shows precisely when assumptions about moment relationships hold, and does so without expanding the number of differential equations, relative to the standard SIR model.
We have performed this integration for the initial distributions shown in Figure~\ref{fig:distribution-evolution}, and the results are shown in Figure~\ref{fig:general-order}A.
Here, we see an orderly progression among the gamma distributions, with the longer-tailed distributions exhibiting reduced final outbreak size; further, we see that the gamma and beta distributions whose initial shape is nearly indistinguishable (gamma, $k=0.25$ with beta, $a=0.25, b=35$) behave almost indistinguishably.
As expected, the gamma distributions follow the power-law relations exactly.
As is apparent, the longer the tail on the distribution, the more suppressed the size of the outbreak, as the initially high rate is caused by those with abnormally high susceptibilities.

While we have only analytically derived a power law for gamma distributions, relating \kt\ to $\cvirgin^\power$, for other distributions we can define an instantaneous order parameter \order\ as

\[ \order \equiv \frac{d (\ln \kt)}{d (\ln \cvirgin) } \]

\noindent
which allows us to understand the deviation from such a law, and can inform us how the contagion process sculpts the susceptibility distribution.
For gamma distributions, $\order = 1/k$ at all times.
We provide an expression for \order\ in terms of \progress\ for the beta distribution in the SI, and in Figure~\ref{fig:general-order}B we show how \order\ changes for the beta distributions.
We see that instantaneous orders of the beta distributions tend towards the orders of the corresponding gamma distributions.
That is, the contagion process sculpts the beta distributions towards gamma distributions, and as this happens the order approaches the constant order of the gamma distribution.
How this affects the outbreak behavior depends upon the relevant timescales of the sculpting process versus that of the infection dynamics.

For the beta distribution at $a=\frac{1}{4}$, $b=35$, which very closely resembles a gamma distribution, we see that the order starts at approximately 4 and closely follows the behavior of the corresponding ($k{=}4$) gamma distribution.
This was also apparent from the numerical integration of these two starting distributions, which give nearly indistinguishable results.
That is, if the initial distribution can be reasonably approximated by a gamma distribution, then we can expect the power-law behavior to describe well the dynamics.
For the more extreme initial distributions, such as the bimodal or uniform, we see that \order\ changes much more dramatically over the course of the outbreak.
Although they approach a gamma distribution at large values of \progress, in practice these \progress\ values may never be reached, as the outbreak may reach its final size before the sculpting process can asymptote.
We see precisely this for the bimodal and uniform distributions, where the dot on the \order\ figures shows the value of \order\ at a time when the epidemic has, in effect, concluded.
Finally, we see the unimodal distribution has a relatively low order parameter, which we expect, as this most closely resembles the implied homogeneous assumption of the classic \sir\ model, for which $\order = 0$ at all times.

How does a power-law dependence of the infection rate affect epidemic model predictions?
First, we note that in the early, exponential-like growth phase of an outbreak---when most of the pre-emptive measures are decided---the models are indistinguishable (since $\cvirgin{\approx}1$, then $\cvirgin^\power{\approx}1$).
This means that when the model parameters are fit in this exponential-like growth phase, the estimate of \ki\ (or alternatively $\Ro \equiv \ki/\kd$) will be identical in each model (see Figure~\ref{fig:h1n1}A).
A deviation in the dynamics of the different models appears given sufficient depletion of susceptibles (see Figure~\ref{fig:h1n1}B).
Assuming the epidemic proceeds in a na{\"i}ve population, the final size under a 2nd-order model is predicted to be

\[ Z = 1 - \frac{1}{\Ro} \hspace{2em} \text{(2nd order)} . \]
\noindent
For 1st- and $p$th-orders, implicit algebraic equations are given in the SI.
Figure~\ref{fig:h1n1}B shows that there can be very substantial differences between final size predictions when comparing the conventional \sir{} model with predictions from models with intrinsic variation in susceptibility.
We note that these results agree with predictions made elsewhere that heterogeneity reduces the predicted severity of the outbreak~\cite{Eksin2019, Gomes2020, Britton2020, Dufresne2020}.
Within the power-law model, the herd-immunity threshold is predicted to be
\[ \hit  =  1 - \frac{1}{\Ro^{1/\power}}. \]
\noindent
These herd-immunity and final-size analyses are only meaningful when the population's immunity is accrued naturally through the dynamics of the outbreak itself; if immunity is acquired through vaccinations applied randomly, then the final-size and herd-immunity estimates will tend toward the classic \sir\ predictions.

Differences in final-size predictions when including risk heterogeneity as compared to those from conventional \sir{} models can be relevant to real-world outbreaks, as we examine for the 2009 H1N1 influenza outbreak, where initial estimates of the final outbreak size were reportedly much higher than occurred in practice~\cite{Nishiura2011}.
In Figure~\ref{fig:h1n1}B, we compare the power-law behavior using models parameterized with the mean \Ro\ estimated at the early stages of the outbreak~\cite{Nishiura2011}.
In the included bar chart (Figure~\ref{fig:h1n1}C), we compare model predictions using 78 published estimates of \Ro\ (from different locales)~\cite{Biggerstaff2014} to final-size measurements based on 11 serological studies~\cite{Nishiura2011}.
We observe that the higher-order models are more consistent with the final-size measurements.
It is not uncommon for models to over-predict the ultimate size of epidemics~\cite{Butler2014,King2015,Nishiura2011}, although we note that non-pharmaceutical interventions such as social distancing are expected to reduce this final size~\cite{Ma2006,Bjornstad2020}, even after being later relaxed.
While predicting final outbreak sizes based on initial measurements is challenging, the significant improvement of the higher-order models lends credence to the importance of accounting for heterogeneity in the susceptibility of na{\"i}ve populations.

There are other factors worth considering in this analysis. 
First, this framework does not assume any correlation between spreading propensity and infection susceptibility, so that super-spreaders appear with a frequency identical to their prevalence in the population. 
It seems plausible that many of the social traits that contribute to higher susceptibility would also lead to higher spreading.
If a positive correlation exists between spreading and susceptibility, then the ``super-spreaders'' will be most active at the beginning of an outbreak and contribute disproportionately to the rate of infection, but as the outbreak progresses, they also will be disproportionately removed early, causing more downward pressure on the relative rate of infection. 
The analysis also assumes that the susceptibility of each individual in the population is static over the timescales relevant to the outbreak, so that an individual does not become more or less susceptible at any point, and that there is no societal response.
If these changes to susceptibility are random, this would decrease the effect suggested here; however, if they are behavior responses, it may increase the effect~\cite{Eksin2019}.

Overall, this study outlines that quantifying and accounting for variability in susceptibility are critical to designing effective policies.
When we account for heterogeneity in infection susceptibility, the rate equation of the standard \sir\ model follows a generic power law with an exponent that increases with the variance and tends toward second-order behavior for an exponential susceptibility distribution.
Other starting distributions can be integrated with negligible computational cost or complexity by relating \kt\ to \progress.
This provides a principled route to account for variability in human populations that are implicitly neglected in classic epidemiological models.
The new framework adds negligible computational cost, facilitating integration with more complex models.
More critically, the analysis reveals that assessing the joint dynamics of the susceptibility distribution can have drastic consequences on naturally-acquired herd immunity predictions, highlighting the importance of quantifying susceptibility variance for COVID-19 and other infectious diseases.

\bibliographystyle{jpclp}
\bibliography{second-order}
\end{document}


\maketitle

\tableofcontents

\section{Emergence of higher-order kinetics}

\subsection{General rate form}

Here, we provide a more detailed derivation of the emergence of higher-order kinetics.
We follow the classic \sir\ model, where the population is divided into susceptible, infectious, and recovered, and define \nvirgin, \nactive, and \ndone\ as the number of people in each of these compartments.
The fraction of the population in these compartments is \cvirgin, \cactive, and \cdone, defined by dividing each of these numbers by the total population, $\ntotal\ = \nvirgin + \nactive + \ndone$.
We also assume a initial epidemic~\cite{Bjornstad2020}, in which the entire population is in principle susceptible to infection (minus the tiny fraction infected at the start of the model).
We examine the initial \sir\ model for simplicity and transparency, and this general framework should lend itself to adaptation to more complex models.

The differential equations that describe the classic \sir\ model can be expressed as the coupled ordinary differential equations:

\begin{equation}\label{eq:classic-sir} \frac{d}{dt} \left[ \begin{array}{c} \cvirgin \\ \cactive \\ \cdone \end{array} \right] = \left[ \begin{array}{l} - \ri \\ \ri - \rr \\ \rr \end{array} \right]
\end{equation}

\noindent
where \ri\ and \rr\ are the rates of those becoming infected and recovering.
Any of the three differential equations can be replaced with $1 = \cvirgin + \cactive + \cdone$.
In all cases, we will keep \rr\ in its standard form; that is, $\rr = \kd \cactive$.

We will examine how the rate of new infections is affected if there is risk heterogeneity in the population.
We define an individual's infection susceptibility \e\ to be proportional to their instantaneous rate of becoming infected.
If we assume the susceptibility is continuously distributed with a distribution given by \dist, then the rate will formally be defined as

\begin{equation}\label{eq:rate-continuous} \rdist =  \ki \, \cactive \, \e \, \frac{\dist}{\ntotal} \end{equation}

\noindent
\rdist\ is properly a distribution of rates; a finite rate for a specific range of susceptibilities is obtained by integrating \rdist\ between any two values of \e.
The total rate of new infections at any point in time is then

\[ \ri = \int_0^\infty \ki \, \cactive \, \e \, \frac{\dist}{\ntotal} \, d\e  \]

\noindent
We can note that the average susceptibility of the susceptible pool can be calculated, at any time, by

\[ \kt = \frac{\int_0^\infty \e \, \dist \, d\e  }{\nvirgin} \]

\noindent
Thus, the rate of new infections simplifies to

\begin{equation}\label{eq:rate-from-mean} \boxed{\ri = \ki \, \cactive \, \kt \, \cvirgin}  \end{equation}

\noindent
where we emphasize that the mean susceptibility \kt\ of this population can change over the course of the epidemic, consistent with the definitions of \kt\ above.

Equation~\eqref{eq:rate-from-mean} is general and applies to any distribution---discrete, continuous, or single-valued.
(For completeness, we also derive this assuming discrete susceptibilities in Section~\ref{sxn:discrete-rate}.)
We also enforce that the average susceptibility is equal to one at the beginning of an outbreak, \kto=1, to ensure compatibility of the definition of \kt\ with the classic model.
(That is, we recover the classic \sir\ model by assuming that every individual is identically susceptible to infection; that is, when $\kt(t) = \kto = 1$.)
We next need to find how this value changes in response to the epidemic.

\subsection{Exponentially-distributed susceptibilities}

For reasons we discuss in the text and in Section~\ref{sxn:distributions}, a natural starting point is to choose a continuous, exponential distribution to describe the variation in infection susceptibility within the population.
This distribution is given by

\[ \dist = \frac{\nvirgin}{\kt} \cdot e^{-\e/\kt} \]

\noindent
We can verify that the total number of individuals in the distribution is \nvirgin\ and that the mean of this distribution is given by \kt:

\[ \int_0^\infty \dist \, d\e = \frac{\nvirgin}{\kt} \int_0^\infty e^{-\e/\kt} \, d\e = \nvirgin \]

\[ \langle \e \rangle = \frac{\int_0^\infty \e \cdot e^{-\e/\kt} d\e}{\int_0^\infty e^{-\e/\kt} d\e} = \frac{ \left[ - \left(\e \kt + \kt^2\right) e^{-\e / \kt} \right]_0^\infty}{ \left[ - \kt e^{-\e / \kt} \right]_0^\infty } = \kt \]

\noindent
For convenience, we will define the total susceptibility of the population as $E \equiv \nvirgin \cdot \kt = \int_0^\infty \e \, \dist \, d\e$.
The average \e\ of a person removed from the population, at any instant in time, is found from the distribution of rates as (see note in Section~\ref{sxn:SI-mean-e} below)

\begin{equation} \label{eq:e-removed}
	\langle \e \rangle_\mathrm{removed} \equiv \frac{\int_0^\infty \e \cdot \rdist \, d\e}{\int_0^\infty \rdist\, d\e} =
\frac{\left[ - \kt \cdot \left(\e^2 + 2 \kt \, \e + 2 (\kt)^2 \right) e^{-\e/\kt} \right]_0^\infty}{\left[ - \kt \cdot \left( \e + \kt \right) e^{-\e/\kt} \right]_0^\infty} = 2 \kt \end{equation}

\noindent
That is, the average not-yet-infected person has susceptibility of \kt, while the average person becoming infected has susceptibility $2 \kt$.
This is valid at any time, provided the population stays exponentially distributed (which, as we discuss in Section~\ref{sxn:distributions} is the expected behavior if the initial distribution is exponential).
Therefore, the total susceptibility of the pool changes as

\[ \frac{dE}{d\nvirgin} = 2 \kt \]

\noindent
Since $E = \kt \, \nvirgin$, it follows that $\frac{dE}{d\nvirgin} = \kt + \nvirgin \frac{d \kt}{d\nvirgin}$. Combined with the previous result, we obtain:

\[ \frac{d \kt}{d\nvirgin} = \frac{\kt}{\nvirgin} \]

\noindent
which upon integration gives 

\[ \frac{\kt}{\kto} = \frac{\nvirgin}{\nvirgino} \]

\noindent
For the case of an initial outbreak $\kto = 1$ and $\nvirgino \approx \ntotal$, so we find

\begin{equation}\label{eq:kt-cvirgin} \boxed{\kt = \cvirgin} \end{equation}

\noindent
That is, the average susceptibility decreases in direct proportion to the fraction of the population still susceptible.
Inserting this into our rate equation, \eqref{eq:rate-from-mean}, gives rise to second-order kinetics:

\[ \boxed{\ri = \ki \cactive \cvirgin^2 }\]

\noindent
And thus, if we account for the heterogeneity of infection susceptibility in the population with an exponential distribution, the \sir\ model should be written as

\[  \frac{d}{dt} \left[ \begin{array}{c} \cvirgin \\ \cactive \\ \cdone \end{array} \right] = 
\left[ \begin{array}{l} - \ki \cactive \cvirgin^2 \\
	\ki \cactive \cvirgin^2 - \kd \cactive \\
\kd \cactive \end{array} \right] \]

\noindent

\subsection{Changes to the mean susceptibility with time}

The generalization of equation~\eqref{eq:e-removed} for any distribution \dist\ is

\begin{equation}
\langle \e \rangle_\mathrm{removed} \equiv
\frac{\int_0^\infty \e \cdot \rdist \, d\e}{\int_0^\infty \rdist\, d\e} =
\frac{\int_0^\infty \e^2 \, \dist \, d\e}{\int_0^\infty \e \, \dist\, d\e} =
\frac{\sigma^2 + \kt^2}{\kt} =
\frac{\sigma^2}{\kt} + \kt
\end{equation}
\noindent
where $\sigma^2$ is the variance, which for a continuous function can be written as $\sigma^2 = \int \e^2 \dist d\e - \kt^2$.
That is, this gives the average susceptibility of those removed from the pool at any time $t$.
By identical manipulations as in the exponential case, we can find a differential equation for how the mean susceptibility changes with the pool size:

\[ \frac{d\left( \kt \, \nvirgin \right)}{d \nvirgin} = \kt + \nvirgin \frac{d \kt}{d\nvirgin} = \frac{\sigma^2}{\kt} + \kt  \]

\begin{equation}\label{eq:mean-evolution} \boxed{\frac{d \kt}{d\nvirgin} = \frac{\sigma^2}{\kt \, \nvirgin}} \text{ or }
 \boxed{\frac{d \kt}{d\cvirgin} = \frac{\sigma^2}{\kt \, \cvirgin}} \end{equation}

 \noindent
 since $\cvirgin(t) = \nvirgin(t)/\ntotal$.

\subsection{Gamma-distributed susceptibilities}

The gamma distribution, at mean \kt\ and with shape parameter $k$, is written as

\[ \frac{\dist}{\nvirgin} = \frac{\e^{k-1} \, e^{- \e k / \kt}}{ \left( \frac{\kt}{k} \right)^k \, \Gamma(k)} \]

\noindent
As we show in Section~\ref{sxn:gamma-evolution}, the gamma distribution is an ``eigendistribution'' of these dynamics; that is, if the initial susceptibility is gamma distributed with shape $k$, the distribution at all times will remain gamma-distributed with shape $k$ (but in general with changing mean).

The variance of the gamma distribution is $\sigma^2 = \kt^2/k$, which if we use with equation~\eqref{eq:mean-evolution} leads to

\[ \int_1^{\kt} \frac{d \, \kt}{\kt} = \frac{1}{k} \int_{1}^{\cvirgin} \frac{d \cvirgin}{\cvirgin} \]

\[ \kt = \cvirgin^{1/k} \]

\noindent
(For an initial outbreak with naturally acquired infections.)
Thus, the rate equation for gamma-distributed susceptibilities becomes a general power-law form based on the shape parameter $k$ of the gamma distribution:

\[ \boxed{\ri = \ki \cactive \cvirgin^{1 + \frac{1}{k}} }\]

We see that when $k{=}1$, we recover the second-order behavior of the exponential distribution, and when $k{\to}\infty$ we recover the first-order model of an identically-susceptible population.
The differential equations of the \sir\ model with gamma-distributed susceptibility is therefore:

\[  \frac{d}{dt} \left[ \begin{array}{c} \cvirgin \\ \cactive \\ \cdone \end{array} \right] = 
\left[ \begin{array}{l} - \ki \cactive \cvirgin^\power \\
	\ki \cactive \cvirgin^\power - \kd \cactive \\
\kd \cactive \end{array} \right] \]

\noindent
where $\power = 1 + 1/k$.

\subsection{Note: instantaneous average susceptibility of those withdrawn \label{sxn:SI-mean-e}}

Here, we provide justification for equation~\eqref{eq:e-removed}.
(Note that for simplicity, we referred to this as the ``draw probability'' in the main text; what we provide here is more rigorous.)
Let's first consider the simple case where we just have two discrete susceptibilities, $\e_1$ and $\e_2$.
The average \e\ of those individuals withdrawn (infected) in the time period between $t$ and $t + \delta t$ is

\[ \langle \e \rangle_\mathrm{removed} = \frac
{ \e_1 \cdot \left( \parbox{9em}{\raggedright number withdrawn with $\e_1$ in $(t, t + \delta t)$} \right)
+ \e_2 \cdot \left( \parbox{9em}{\raggedright number withdrawn with $\e_2$ in $(t, t + \delta t)$} \right) }
{\left( \parbox{9em}{\raggedright number withdrawn with $\e_1$ in $(t, t + \delta t)$} \right) 
+ \left( \parbox{9em}{\raggedright number withdrawn with $\e_2$ in $(t, t + \delta t)$} \right)}
\]

\[ \langle \e \rangle_\mathrm{removed} = \frac
{\e_1 \cdot \int_t^{t+\delta t} r_1 \, dt 
+ \e_2 \cdot \int_t^{t+\delta t} r_2 \, dt }
{\int_t^{t+\delta t} r_1 \, dt 
+ \int_t^{t+\delta t} r_2 \, dt }
\]

\noindent
Taking the limit as $\delta t \to 0$ gives us the instantaneous average \e\ of those withdrawn at time $t$ (using L'H{\^o}pital's Rule and the Fundamental Theorem of Calculus):

\[ \langle \e \rangle_\mathrm{removed} = \frac{\e_1 r_1 + \e_2 r_2}{r_1 + r_2} \]

\noindent
This can be generalized to any number of levels as

\[ \langle \e \rangle_\mathrm{removed} = \frac{\sum_i \e_i r_i}{\sum_i r_i} \]

\noindent
In the continuum limit, the above becomes

\[ \langle \e \rangle_\mathrm{removed} = \frac{\int_0^\infty \e \cdot \rdist \, d \e}{\int_0^\infty \rdist \, d\e} \]

\subsection{Note: discrete susceptibility levels \label{sxn:discrete-rate}}

Here, we show that equation~\eqref{eq:rate-from-mean}---that is $ \ri = \ki \kt  \cactive  \cvirgin$---also arises if we assume that the susceptibility is binned into discrete levels.
If we have discrete levels of susceptibility in a population, a rate equation of new infections is written for each value of susceptibility \ei\ as

\[ r(\ei) = \ki \, \cactive \, \ei \, \cvirgini \]

\noindent
where \cvirgini\ is the fraction of the total population with susceptibility \ei.
The total rate of new infections at any point in time is then

\[ \ri = \sum_i \ki \, \cactive \, \ei \, \cvirgini  \]

\noindent
The average susceptibility of the susceptible pool can be calculated, at any time, by

\[ \kt = \sum_i \ei \cvirgini  \]

\noindent
Thus, the rate of new infections again simplifies to

\[ \ri = \ki \, \cactive \, \kt \, \cvirgin \]

\section{Instantaneous order parameter}

In the main text, we defined an instantaneous order parameter that allows us to compare an arbitrary distribution to a power-law, as

\[ \order \equiv \frac{d (\ln \kt)}{d (\ln \cvirgin)} \]

\noindent
That is, for a distribution that follows a power law (specifically, the gamma), \order\ will be a constant at all times during the pandemic.
We provide the value for \order\ for the various distributions in Section~\ref{sxn:distributions}.
Here, we show that $\order = \sigma^2 / \kt^2$; that is, the instantaneous order parameter is the square of the coefficient of variation at any time.

We can find this by starting with equation~\eqref{eq:mean-evolution}

\[ \frac{d \kt}{d\cvirgin} = \frac{\sigma^2}{\kt \, \cvirgin} \]

\noindent
Re-arranging, and substituting $dx/x = d(\ln x)$,

\[ \frac{d \cvirgin}{\cvirgin} = \frac{\kt}{\sigma^2} d \kt \]

\[ d (\ln \cvirgin) = \frac{\kt^2}{\sigma^2} d (\ln \kt) \]

\begin{equation}\label{eq:order} \order \equiv \frac{d (\ln \kt)}{d (\ln \cvirgin)} = \frac{\sigma^2}{\kt^2} \end{equation}

\section{How distributions are sculpted by contagion\label{sxn:distributions}}

\subsection{General sculpting relations}

Here, we derive how distributions change shape over time, which in general will show that gamma-distributed functions (including the exponential distribution) will remain in the same shape through the action of contagion.
Consider a population of susceptible individuals with a distribution given by \dist.
The distribution will change in time according to the partial differential equation:

\[ \frac{\partial \, \dist}{\partial t} = - \ki \e \cactive \dist \]

\noindent
Integrating at a particular value of \e\ gives

\[ \dist = \disto \cdot \exp \left\{ -\ki \e \int_0^t \cactive dt \right\} \]

\begin{equation}\label{eq:evolution} \boxed{ \dist = \disto \cdot e^{-\progress \, \e } } \end{equation}

\noindent
where we have defined $\progress \equiv \ki \int_0^t \cactive \, dt$ as a dimensionless progress variable that is monotonic with time (since $\cactive \ge 0$).
This equation shows how any distribution will change over time under the force of contagion.

\subsection{Examples with specific classes of distributions}

Next, we show how the sculpting process affects selected initial distributions.

\subsubsection{Exponential distribution}

If the original susceptibility distribution is exponential with initial mean \kto; that is,

\[ \disto = \frac{\ntotal}{\kto} e^{-\e/\kto}  \]

\noindent
then equation~\eqref{eq:evolution} gives

\[ \dist = \frac{\ntotal}{\kto} \exp \left\{-\left(\frac{1}{\kto} + \progress \right) \cdot \e \right\} \]

\noindent
That is, the distribution will remain exponentially distributed, with respect to \e, at all times.
By inspection of the form, we can see that the mean changes as

\[ \kt(t) = \frac{1}{\frac{1}{\kto} + \progress } \]

\noindent
The variance for an exponential distribution is $\sigma^2 = \kt^2$, and since the distribution stays exponentially distributed at all times, the order parameter will therefore be $\order = 1$ at all times, by equation~\eqref{eq:order}.
(This was also obvious by inspection of equation~\eqref{eq:kt-cvirgin}.)

\subsubsection{Gamma distribution\label{sxn:gamma-evolution}}

If the original susceptibility distribution is gamma-distributed with initial mean \kto\ and shape parameter $k$; that is,

\[ \disto = \ntotal \, \frac{\e^{k-1} e^{-\e k / \kto} }{ \left( \frac{\kto}{k} \right)^k \Gamma(k) } \]

\noindent
(where $\Gamma(k)$ is the gamma function) then equation~\eqref{eq:evolution} gives

\[ \dist = \ntotal \, \frac{\e^{k-1} e^{-\left( \frac{k}{\kto} + \progress \right) \e} }{ \left( \frac{\kto}{k} \right)^k \Gamma(k) } \]

\noindent
which is itself a gamma distribution, which by inspection has shape $k$ and mean

\[ \kt(t) = \frac{1}{ \frac{1}{\kto} + \frac{\progress}{k}} \]

\noindent
That is, a gamma distribution of shape $k$ stays a gamma distribution of shape $k$ under the action of contagion.
Thus, we can consider this an ``eigendistribution''.
Note that the exponential distribution can be expressed as a gamma distribution with shape $k=1$.

For the gamma distribution, the variance is $\sigma^2 = \kt^2 / k$.
Since the gamma distribution retains its shape, by equation~\eqref{eq:order} the instantaneous order is $\order = 1/k$ at all times.

\subsubsection{Beta distribution}

If the original susceptibility distribution is beta-distributed with initial mean \kto\ and shape parameters $a$ and $b$; that is,

\[ \disto = \ntotal \, \frac{\left(\frac{\e}{\emax}\right)^{a - 1} \left(1 - \frac{\e}{\emax}\right)^{b-1}}{B(a,b)} \]

\noindent
where $B(a,b) = \Gamma(a) \Gamma(b) / \Gamma(a + b)$ is a normalizing factor.
This distribution has been scaled to have initial mean \kto\ and is supported on $(0,\emax)$, with $ \emax/\kto = (a + b) /a$.
This distribution changes over time as

\[ \disto = \ntotal \, \frac{\left(\frac{\e}{\emax}\right)^{a - 1} \left(1 - \frac{\e}{\emax}\right)^{b-1} e^{-\progress \e}}{B(a,b)} . \]

\noindent
The mean time course of this distribution can be found to be

\[ \kt(t) = a \emax \frac{\reghyp(a + 1, a + b + 1, - \emax \progress)}{\reghyp(a, a+b, -\emax \phi)} \]

\noindent
where \reghyp\ is the regularized confluent hypergeometric function.
(Note that \reghyp\ is available as a function call in most computational packages.)
The instantaneous order parameter can be found from equation~\eqref{eq:order} to be:

\[ \order = \frac{a+1}{a} \cdot \frac{\reghyp(a, a+b, -\emax \progress) \cdot \reghyp(a+2, a+b+2, -\emax \phi)}{\left(\reghyp(a+1, a+b+1, -\emax \progress)\right)^2} - 1 \]

\subsection{Which distributions tend towards gamma?}

In the main text, we show that beta distributions---even with rather extreme-looking initial features---are gradually sculpted towards gamma distributions through the process of contagion, where the shape $k$ of the emerging gamma distribution was directly predicted from one of the two shape parameters ($a$) of the beta distribution.
Meanwhile, gamma distributions stay perfectly gamma-distributed, maintaining their shape factor $k$ as the mean susceptibility decreases.
How universal is this behavior---of distributions tending towards gamma?

Here, we put together mathematical criteria for when we expect that a gamma distribution will \emph{eventually} emerge through the sculpting process---as well as what shape parameter $k$ that distribution will have---but again caution that these limiting values (written in terms of the cumulative infection progress variable \progress) may not in practice be approached, if the timescale of reaching the outbreak final size is short compared to the timescale of the sculpting process.
We will also treat these distributions as continuous and infinitely-divisible, ignoring the discrete nature of human beings.

If \disto\ is continuous and can be expressed as a power series, then we can show that \dist\ will approach a gamma distribution as $\progressnot \to \infty$.
We will examine some region of susceptibilities on $\eminus \le \e < \eplus$, where $\eminus$ approaches zero.
(Note that the gamma distribution is only defined on $\e>0$.)
If \disto\ can be expressed as a power series about \eminus\ over this domain, we can expect that---for a narrow enough domain---the leading term of the power series will dominate.
Let's say that leading term is $\theta \cdot (\e - \eminus)^l$, where $\theta$ and $l$ are constants with $l>-1$ and $\theta>0$; in this case, by equation~\eqref{eq:evolution}, the approximate distribution evolves as:

\[ \dist \approx \theta \cdot (\e - \eminus)^l \cdot e^{-\progress \cdot \e} \text{ ,} \]

\noindent
which is a gamma distribution with shape $k=l+1$, as $\eminus \to 0$.
Since $e^{-\progress \cdot \e}$ concentrates the probability density to lower values of \e\ as \progress\ increases, even if \disto\ takes on arbitrarily strange shapes at high values of \e, so long as a small region can be found where \disto\ is well-approximated by the first term of its power series, then the gamma behavior will eventually be seen at large values of \progress.

\subsubsection{Finding the appropriate shape $k$ for the beta distribution}

If the susceptibilities are initially beta-distributed, then the population's probability density is initially proportional to $f(\e,0) = \e^{a-1} (1-\e)^{b-1}$.
The power series representation of $f(\e,0)$ is

\[ f(\e,0) = \sum_{n=0}^{\infty} \left[ \frac{(-1)^n}{n!} x^{n + a - 1} \prod_{m=0}^{n-1} (b - 1 - m) \right] \]

\noindent
The leading term ($n=0$) of this series is $(b-1) x^{a-1}$, and thus in the limits above the probability density evolves with a shape approaching

\[ f(\e, t) \approx (b-1) \cdot (\e - \eminus)^{a-1} \cdot e^{-\progress \cdot \e} \]

\noindent
which as $\eminus \to 0$ is a gamma distribution with shape $k=a$.
This agrees with the results in Figures 2 and 3B of the main text, in which we observed initially-beta distributions tend towards the shape and apparent-order of analogous gamma distributions with shape $k=a$.

\section{Running \sir-type models with arbitrary initial distributions}

Equation~\eqref{eq:rate-from-mean} told us that the rate of new infections, for any distribution, can be expressed as $\ri = \ki \, \cactive \, \kt \, \cvirgin$, where \kt\ is the (instantaneous) mean susceptibility of the pool of susceptibles.
In Section~\ref{sxn:distributions} we showed how \kt\ can be related to $\progress \equiv \ki \int_0^t \cactive \, dt$ for arbitrary distributions, and gave results for the exponential, the gamma, and the beta distributions.
(Although we note that a simple power law works for the exponential and gamma distributions, where the current procedure would be unnecessary.)
Therefore, if we can track \progress\ over the course of the outbreak, we have the information to construct equation~\eqref{eq:rate-from-mean}.

Since in the \sir\ model, $\cdone(t) = \kd \int_0^t \cactive \, dt$, then at any time $\progress = \frac{\ki}{\kd} \cdone(t)$.
Therefore, \progress\ can be tracked with nearly no extra complexity or computational cost, since the value of $\cdone(t)$ is always known.

Thus, we can model any arbitrary starting distribution (for which we can express $\kt(\phi)$) using only two coupled ordinary differential equations, the same system size as the classic \sir\ model.
Of course, these quantities may need to be re-derived for alternate models, such as SEIR.
We note also that \progress\ could alternatively be tracked by noting that $d \progressnot / dt = \ki I$, where this differential equation could be explicitly added to the system that is numerically integrated.

\subsection{Example: Beta distribution}
As an example, to run the \sir\ model with an initially beta distribution, one would numerically integrate the system

\[ \frac{d}{dt} \left[ \begin{array}{c} \cvirgin \\ \cactive \end{array} \right] = \left[ \begin{array}{l} - \ri \\ \ri - \rr \end{array} \right] \]

\noindent
where the values of \ri\ and \rr\ at any time are found from the explicit algebraic equations:

\[ \progress = \frac{\ki}{\kd} \cdot \left( 1 - \cvirgin(t) - \cactive(t) \right) \]

\[ \kt(t) = a \,  \emax \frac{\reghyp(a + 1, a + b + 1, - \emax \, \progress)}{\reghyp(a, a+b, -\emax \, \phi)} \]

\[ \ri = \ki \, \cactive(t) \, \kt(t) \, \cvirgin(t) \]

\[ \rr = \kd \, \cactive(t) \]

\section{Herd immunity thresholds and final-size calculations}

We emphasize that both of the results below are valid when immunity is obtained naturally; if immunity is instead obtained randomly (as may be a more appropriate assumption if immunizations are employed) then the first-order model may be a more appropriate assumption.

\subsection{Herd immunity threshold}

The herd immunity threshold is defined as the fraction of the population that must be infected for the rate of new infections to go into decline.
This occurs when $d\cactive/dt=0$ (that is, when the sign changes from positive to negative); which in the (power-law) \sir\ model is

\[ \frac{d\cactive}{dt} = +\ki \cactive \cvirgin^\power - \kd \cactive = 0 \]

\noindent
where $\power = 1 + 1/k$ for the k-shaped gamma distribution.
That is, it occurs when

\[ S = \left(\frac{\gamma}{\beta}\right)^{1/\power} = \frac{1}{\Ro^{k/(k+1)}} \]

\noindent
where $\Ro \equiv \ki/\kd$.
The fraction of the population that must be infected is often denoted as \hit, and is therefore

\[ \hit  = 1 - \Ro^{-k/(k+1)} \]

\subsection{Final-size calculations}

The predicted final size of an outbreak is the proportion of the population that is ever infected if the outbreak is allowed to proceed without intervention;~\cite{Kermack1927,Ma2006} however, we note that even temporary interventions can affect the final size prediction.
(That is, if \Ro\ is temporarily reduced through non-pharmaceutical interventions that are later relaxed.)
The final size is traditionally found by forming $d\cactive/d\cvirgin$:

\[ \frac{d \cactive}{d\cvirgin} = \frac{d\cactive}{dt} \cdot \frac{dt}{d\cvirgin}
=  \frac{\ki \cactive \cvirgin^\power - \kd \cactive}{-\ki \cactive \cvirgin^\power}
= \frac{1}{\Ro \cvirgin^\power} - 1 \]

\noindent where $\Ro \equiv \ki/\kd$ and $\power = 1 + 1/k$ for the $k$-shaped gamma distribution.
This is integrated from the beginning to the end of the outbreak:

\[ \int_{\cactiveo}^{\cactivef} d\cactive = \int_{\cvirgino}^{\cvirginf} \left( \frac{1}{\Ro \cvirgin^\power} - 1 \right) d \cvirgin \]

\noindent
where the subscripts 0 and $\infty$ mean the quantities are evaluated at $t{=}0$ and $t{\to}\infty$, respectively.
The outbreak ends with $\cactivef = 0$; also employing $\cactiveo\approx0$ and $\cvirgino\approx1$ as bounds we find

\[ 0 = \frac{1}{1-\power} \left( \frac{1}{\Ro \cvirginf^{\power-1}} - \frac{1}{\Ro} \right) - \cvirginf + 1 \hspace{3em} \text{($\power$th-order)}  \]

\[ 0 = \frac{1}{\Ro} \ln \cvirginf - \cvirginf + 1 \hspace{3em} \text{(1st-order)} \]

\noindent
A closed-form solution is available when $\power=2$.
In other cases an implicit algebraic formula is provided.
The final size is defined as $\finalsize \equiv 1 - \cvirginf$ giving:

\[ \finalsize = 1 - e^{-\finalsize / \Ro} \hspace{3em} \text{(1st-order)} \]

\[ \finalsize = 1 - \frac{1}{\Ro} \hspace{3em} \text{(2nd-order)} \]

\[ \finalsize = \frac{1}{(1 - \power) \Ro} \left( 1 - \frac{1}{(1 - Z)^{\power - 1}} \right) \hspace{3em} \text{($\power$th-order)} \]

\bibliographystyle{jpclp}
\bibliography{../second-order.bib}